\documentclass[a4paper]{jpconf}
\usepackage{graphicx}
\begin{document}
\title{CP violation and final state interactions in $B\to K\pi\pi $ decays}

\author{L. Le\'sniak$^1$, B. El-Bennich$^2$, A. Furman$^3$, R. Kami\'nski$^1$,
B. Loiseau$^2$ and B. Moussallam$^4$}

\address{$^1$ The Henryk Niewodnicza\'nski Institute of Nuclear Physics, Polish 
Academy of Sciences,\\ 31-342 Krak\'ow, Poland}
\address{$^2$ Laboratoire de Physique Nucl\'eaire et de Hautes \'Energies 
(IN2P3-CNRS-Universit\'es Paris 6 et Paris 7), Groupe Th\'eorie,
Universit\'e Pierre et Marie Curie, 4 Pl. Jussieu,\\ F-75252 Paris, France}
\address{$^3$ ul. Bronowicka 85/26, 30-091 Krak\'ow, Poland}
\address{$^4$ Institut de Physique Nucl\'eaire, Universit\'e Paris-Sud,
F-91406 Orsay, France}

\ead{Leonard.Lesniak@ifj.edu.pl}

\begin{abstract}

Effects of CP violation and of final state interactions between pairs of 
pseudoscalar mesons are studied in three-body $B^+, B^-, B^0$ and 
$\bar B ^0$ decays into $K\pi\pi$.
An alternative approach to the isobar model for three-body B decays is 
proposed. It is based on the QCD factorization approximation and the knowledge 
of the meson-meson form factors. Some
phenomenological charming penguin amplitudes are needed to 
describe the branching fractions, direct CP asymmetries of the quasi-two-body
$B \to K^*(892) \pi$ and $B \to K_0^*(1430) \pi$ decays as well as 
the $K\pi$ effective mass and the helicity
angle distributions. The experimental 
branching fractions for the $B \to K_0^*(1430) \pi$ decay, obtained by the 
Belle and BaBar collaborations using the isobar model, are larger than our
predictions by about 52 per cent.  

\end{abstract}

\section{Introduction}
CP violation in the charged B meson decays has been discovered for the first 
time in the three body charmless hadronic reaction 
$B\to \rho(770)^0 K$, $\rho(770)^0 \to \pi^+\pi^-$ \cite{BB1,Belle2}. Data on
three body $B$ decays are commonly analyzed using the isobar model which 
violates unitarity and requires many free parameters. In \cite{prd} 
we have developed a model with a unitarized description of final state
interactions between the two pions. We found interesting interferences between
the $\rho(770)^0$ and $f_0(980)$ resonances. Within a similar approach and for 
the $K\pi$ effective masses smaller than about 1.8 GeV,
we study another part of the Dalitz plots corresponding to the
$B^{\pm}$, $B^0$ and $\overline{B}^0$ decays where the strange resonances $K^*$ 
are observed. Among 
them, one encounters two resonances $K^*(892)$ and $K^*_0(1430)$ formed in the $K
\pi$ $P$- and $S$- waves, respectively. These two resonances dominate the 
pion-kaon vector and scalar form factors which play an essential role in the 
description of the decay amplitudes. The strange form factors are
determined from unitary coupled-channel equations using the experimental data
on the $\pi K$ phase shifts and inelasticities and some constraints from chiral
perturbation theory. 

\section{Decay amplitudes for $B\to K\pi\pi$ reactions}
The amplitudes for the $B\to K\pi\pi$ decays consist of two terms. The first
part corresponds to the quark weak transitions $b\!\to\! s\overline d d$ or
$b\!\to\! s\overline u u$ in the QCD factorization approximations.  The second 
one is a phenomenological long-distance amplitude with $c-$ or $u-$quarks in 
loop. Such an
amplitude with the $c-$quark in loop, called the charming penguin term \cite{Ciu},
 could
be related to $B$ decay processes with intermediate $D_s^{(*)} D^{(*)}$ states.
For example, the $S$-wave part of the $B^- \to  K^-\pi^+\pi^-$ decay
amplitude reads: 
  
\begin{eqnarray} 
A_S=\frac{G_F}{\sqrt{2}} (M_{B}^2-m_{\pi}^2)\frac{m_{K}^2-m_{\pi}^2}{q^2}
F_0^{B\to\pi}(q^2) f_0^{K^-\pi^+}(q^2)
\times \bigg\{
\lambda_u(a_4^{u}+P_u-a_{10}^{u}/2)+\nonumber\\\lambda_c(a_4^c+P_c-a_{10}^c/2)
 - \frac{2 q^2}{(m_b-m_d)(m_s-m_d)}
\times \left[\lambda_u(a_6^{u}+S_u-a_8^{u}/2)+\lambda_c(a_6^c+S_c-a_8^c/2)
\right]\bigg\} \ .
\end{eqnarray}
Here $q^2$ denotes the  $K^-\pi^+$ effective mass squared, $F_0^{B\to\pi}(q^2)$
is the scalar $B$ to $\pi$ form factor, $f_0^{K^-\pi^+}(q^2)$ is the scalar
$K^-\pi^+$ form factor and the remaining symbols are fully explained in ref.
\cite{prd}. The  $P$-wave amplitude, proportional to the vector form factors
$F_1$ and  $f_1$, is given by   
\begin{equation}
A_P=2\sqrt{2}G_F 
 \ \mathbf{p}_{\pi^-}\cdot\mathbf{p}_{\pi^+}  \ F_1^{B\to\pi}(q^2)
 f_1^{K^-\pi^+}(q^2) 
\left[
 \lambda_u\left(a_4^{u}+P_u-a_{10}^{u}/2\right)+ 
      \lambda_c\left(a_4^c+P_c-a_{10}^c/2\right)
\right] \ .
\end{equation}
The long-distance penguin amplitudes are parametrized by four complex parameters
$S_u, S_c$, $P_u$ and $P_c$.
\section{Comparison with experimental results}
We analyse the experimental data of the Belle and BaBar collaborations 
\cite{BB1,Belle2,Belle1,BB2} for the four reactions: $B^- \to (K^-\pi^+)\pi^- $,
$B^+ \to (K^+\pi^-)\pi^+$, $\overline B^0\to (\overline K^0 \pi^-)\pi^+$ and
$B^0\to (K^0 \pi^+)\pi^-$. The physical observables fitted in our model include
12 values of the branching fractions and direct CP-violating asymmetries for the
 $B \to K^*(892)^0\pi$ and $B \to K_0^*(1430)^0\pi$ decays in different charge
 combinations and 249 data points for the $K \pi$ effective mass and helicity
 angle distributions. If the charming penguin parameters are not included in 
 the fit then the theoretical values of the branching fractions for these decays 
 are 
 too small by factors ranging between 2.3 and 3.6. The results without 
the charming penguin terms are in agreement
 with those recently calculated by Cheng, Chua and Soni \cite{Cheng}. We have
 then
 performed two separate fits with 4 penguin parameters: in the first fit
 the experimental branching fractions for the 
 $B \to K_0^*(1430)^0\pi$ decays have been included while in the second fit they
 have been omitted. The fit was much better in the second case.
 
  Typical results
 of the $K \pi$ effective mass distributions for these two cases are shown in
 Fig. 1(a) and 1(b), respectively. In Fig. 1(a) the theoretical curve
 underestimates the narrow maximum of the $K^*(892)$ resonance and
 overestimates the wide  maximum due to the $K^*_0(1430)$ resonance. On the
 contrary, in Fig. 1(b)  both maxima are well reproduced.
  The average
 branching fractions and asymmetries for the second fit are given in Table 1.
  The CP asymmetries are in a general
 agreement with data. The theoretical value of the 
 $B^{\pm} \to K^*(892)^0\pi^{\pm}$ branching fraction is close to the Belle and
  BaBar results, however
 the corresponding $B^{\pm} \to K_0^*(1430)^0\pi^{\pm}$ value, equal to
  $12.9 \pm 0.6$,
 is substantially smaller than the experimental values. One should stress the
 fact that these experimental numbers are {\it not} directly measured. They are
 calculated by both collaborations using the isobar model. We calculate our
 branching fractions simply by integrating the $K \pi$ effective mass distributions
 over the energy range specified in Table 1. The published experimental 
 branching fractions correspond to the full range of the $K \pi$ effective masses.
 In Table~1, their values are renormalized for the $K^*(892)$ and $K_0^*(1430)$ 
 by 0.83 and 0.81, respectively.  
   The renormalized average branching fractions for the 
   $B^{\pm} \to K_0^*(1430)^0\pi^{\pm}$~decays are about 52 \% larger than the
   result of our model fitted to the same $K \pi$ effective mass and helicity 
   angle data. The 
 published experimental 
 branching fractions are then too large by the same
 amount. A partial explanation of this fact can rely on the presence of a
  phenomenological
  nonresonant part of the decay amplitudes used in the framework of the
 isobar model. In our approach both resonant and nonresonant parts are included
 in the strange $K\pi$ scalar form factor.

\begin{figure}[h]
\begin{center}
\includegraphics[width=32pc]{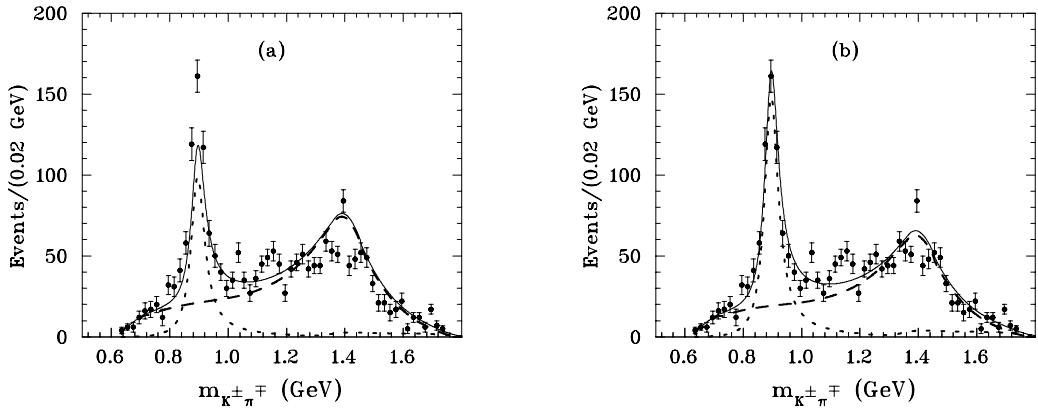}
\end{center}
\caption{\label{label}Averaged $K \pi$ effective mass distributions for the 
$B^{\pm} \!\to\!K^{\pm} \pi^{\mp}\pi^{\pm}$ decays. The dotted line corresponds
to the $P$-wave, the dashed line to the $S$-wave and the solid one to the
full result. The experimental $B \to K_0^*(1430)^0\pi$ values of the branching
fractions are included in the fit shown in (a) but not included in
that shown in (b). Data are taken from \cite{Belle2}. }
\end{figure}
 
\begin{table}[h]
\caption{\label{tabone}Average branching fractions Br$\cdot 10^6$ and direct CP
 asymmetries $A_{CP}\cdot 10^2$ corresponding to the fit shown in Fig. 1(b).} 
 
\begin{center}
\lineup
\begin{tabular}{*{6}{l}} 
\br
obs.   & channel                     & $K\pi$ mass range [GeV]& our model &
~~~~Belle         &~~~ BaBar\cr
\mr
Br           & $K^*(892)^0\pi^{\pm}$& $\0\0\00.82-0.97$&$\0~6.14\pm
0.15$&$\0~~5.35\pm 0.59$&$\0~7.46\pm 0.81$\cr 
$A_{CP}$ & $K^*(892)^0\pi^{\pm}$& $\0\0\00.82-0.97$       & $-8.0\pm 2.4$ 
&$-14.9\pm 6.8$& $\0~6.8\pm
10.4$ \cr
Br           & $K^*_0(1430)^0\pi^{\pm}$& $\0\0\01.00-1.76$& $~12.9\pm 0.6$
&$\0~25.9\pm 2.5$& $~27.5\pm 2.2$\cr 
$A_{CP}$ & $K^*_0(1430)^0\pi^{\pm}$& $\0\0\01.00-1.76$    &$-0.4\pm 1.3$  &$\0~~7.6\pm 4.5$&$-6.4\pm
4.0$ \cr
\br
\end{tabular}
\end{center}
\end{table}
\ack

This work has been performed within the framework of the IN2P3-Polish 
Laboratories Convention (Project No. CSI-12) and within an agreement between 
the CNRS (France) and the Polish Academy of Sciences (Project No. 19481).

\smallskip

\end{document}